\documentstyle[12pt]{article}

\newtheorem{theorem}{Theorem}[section]

\newtheorem{prop}[theorem]{Proposition}
\newtheorem{defin}[theorem]{Definition}
\newtheorem{cor}[theorem]{Corollary}

\makeatletter \@addtoreset{theorem}{subsection} \makeatother
\renewcommand{\thetheorem}{\relax
  \ifnum\value{subsection}=0 \thesection.\arabic{theorem}\relax
  \else \thesubsection.\arabic{theorem}\fi}
\newcounter{step}

\def\(#1){$\rm(#1)$}
\newcounter{Rom}

\newcounter{ara}

\newcounter{itemno}
\newenvironment{items}{\list{\roman{itemno})}{\relax
  \usecounter{itemno}\settowidth{\labelwidth}{\(iii')}\labelsep 0.5em
  \leftmargin\labelwidth\advance\leftmargin by \labelsep
  \def\makelabel##1{##1\hfill}}}{\endlist}

\newcommand{\ie}{i.\,e.~}

\newcommand{\be}{\begin{displaymath}}
\newcommand{\ee}{\end{displaymath}}
\newcommand{\bea}{\begin{eqnarray*}}
\newcommand{\eea}{\end{eqnarray*}}

\newcommand{\Iff}{\mbox{ $\Leftrightarrow$ }}

\newcommand{\Then}{\mbox{ $\Rightarrow$ }}

\newcommand{\fa}{\forall}

\newcommand{\A}{{\cal A}}
\newcommand{\B}{{\cal B}}

\renewcommand{\H}{{\cal H}}

\renewcommand{\phi}{\varphi}
\renewcommand{\epsilon}{\varepsilon}
\renewcommand{\theta}{\vartheta}

\newcommand{\ca}{\mbox{$C^{*}$-algebra}}
\newcommand{\wa}{\mbox{$W^{*}$-algebra}}
\newcommand{\aone}{\mbox{$\A_{1}$}}
\newcommand{\atwo}{\mbox{$\A_{2}$}}
\newcommand{\wotimes}{\overline{\otimes}}
\newcommand{\ten}{\mbox{$\aone\otimes\atwo$}}
\newcommand{\wten}{\mbox{$\aone\wotimes\atwo$}}
\newcommand{\abs}[1]{\mid #1 \mid}
\newcommand{\clet}{Let \aone\ and \atwo\ be \ca s}
\newcommand{\wlet}{Let \aone\ and \atwo\ be \wa s}

\begin{document}
\title{Separation theorems and Bell inequalities
in algebraic quantum mechanics\footnote{This paper appeared in Busch, P., Lahti, P.J.,
Mittelstaedt, P. (eds.): {\em Proceedings of the
Symposium on the Foundations of Modern Physics, Cologne, 1--5 June 1993}, pp. 29--37. Since it is not otherwise available online, I am archiving it now (with minor edits).}}
\author{Guido Bacciagaluppi}
\date{}
\maketitle
\begin{abstract}
The paper discusses the concept of separation of quantum mechanical systems in
the algebraic approach.
We review known theorems, then establish a link
between the $C^*$-algebraic and the corresponding $W^*$-algebraic concepts.
A characterization of separation in terms of Bell inequalities, due to Raggio
(1988), is given a
$C^*$-algebraic formulation. Finally, we comment on the implications for the
understanding of the Bell inequalities.
\end{abstract}

\section{Introduction}
The concept of separation, as we use the term, was introduced by Primas
(1978), and investigated by Raggio (1981, 1983, 1988).
It means the absence of EPR correlations between two subsystems of a
composite system, in the strong sense that the systems are {\em never}
correlated. According to Primas (see for instance his (1983)), the conceptual
relevance of such a concept of separation is that
separated systems are to lack
the holistic features that lead to the EPR paradox, and the subsystems of
a separated system can thus
be treated as individual {\em objects}.

From the point of view of the traditional Hilbert-space formalism this is an
unnatural condition. The states of a product space $\H_1\otimes\H_2$ that lack
EPR-correlations are precisely the product states
$\psi_1\otimes\psi_2$. Requiring that the two systems never be
EPR-correlated is thus a very strong restriction on the dynamics of the system,
indeed, this requirement is satisfied only by noninteracting systems.

However, if one generalises the formalism to algebraic quantum mechanics, this
condition can be met also in the presence of interactions. As an example, take
two concrete von Neumann algebras $\A_1$ and $\A_2$ acting on the Hilbert
spaces $\H_1$ and $\H_2$, and let $\A_1$ be commutative. It is clear that every
element $A$ of $\A_1$ commutes with every element of the product algebra
$\ten$. Intuitively, this means that the possible distributions of the values
of $A$ are not constrained by the uncertainty principle, and so that $A$ has a
dispersion-free value in every pure state $\Psi$ of $\ten$. But now there are
superselection rules separating any two product states
$\psi_1\otimes\psi_2$ and
$\phi_1\otimes\phi_2$ with
$\psi_1\neq\phi_1$. Indeed, if a state of the form
  \be
  \Psi = c_1 \psi_1\otimes\psi_2 +
                     c_2 \phi_1\otimes\phi_2
  \ee
were pure, the reduced state for subsystem 1 would be
  \be
  \rho_1 = \abs{c_1}^2 P_{\psi_1}+\abs{c_2}^2 P_{\phi_1}
  \ee
(where the $P_{\psi}$ are the projectors on the corresponding rays), \ie a
mixture, but then certainly some $A\in\A_1$ would have a non-zero dispersion in
the state $\Psi$, contradicting the above. Thus, if $\A_1$ or $\A_2$ is
commutative, {\em all} pure states are product states, so indeed it is possible
that systems 1 and 2 are never EPR-correlated, {\em independently of the
dynamics}.

Separation was investigated by Raggio (1981, 1983, 1988) in the framework of
$W^*$-algebras. He derived two structure theorems (Th.~\ref{Raggio0} and
\ref{Raggio1} below) that characterise separated
systems completely. In fact, the above example is generic: a $W^*$-system
composed of two subsystems is
separated if and only if
at least one of the subsystems is {\em classical} (\ie if its algebra is
commutative). As Primas (1983) puts it: quantum objects exist
only in a classical environment; or better, idealized models of a quantum
system in which the system is to be separated from its environment must treat
the environment of the system classically.

Later, Raggio seized upon a result of Baez (1987) about
Bell inequalities in the algebraic setting, and
established a remarkable link between Bell inequalities and separation. This is
a particularly interesting aspect of Raggio (1988).

In this paper we examine separation in the framework of \ca s, and in
particular the connection between separation and the Bell inequalities.
In algebraic quantum mechanics, \wa s often arise in connection with
representations of a \ca , and the abstract \ca\ is regarded as more
fundamental.

The definition of a separated $C^*$-system is straightforward, and the
structure theorem analogous to Raggio's theorems is already contained in
Takesaki (1958). In section \ref{results}, we give the definitions and
structure theorems for the $C^*$- and the \wa ic case. We also
make the connection between separation for $C^*$- and $W^*$-systems. In section
\ref{Bell} we turn to the relation between separation and the Bell
inequalities. We prove the converse of Baez's theorem, and
derive the $C^*$-analogue of Raggio's results.
We conclude (section \ref{conc}) with some comments on the significance of
these results for the understanding of the Bell inequalities.

\section{Separation theorems}\label{results}
In the following, when $\A_{1}$ and $\A_{2}$ are $C^{*}$-algebras (which are
always supposed to be unital), \ten\ denotes the injective tensor product
of \aone\ and \atwo . When \aone\ and \atwo\ are \wa s, \wten\ denotes their
$W^{*}$-tensor product.

We now give the exact definitions of separation in the $C^*$- and in the
$W^*$-case.
\begin{defin}
Let \aone , \atwo\ be \ca s (respectively, \wa s). A state (a normalized
positive linear functional) $\omega$ on \ten\ (respectively, \wten ) is said
to be a {\em product state} iff for all $(A_{1},A_{2})$ in $\aone\times\atwo$,
  \be
  \omega(A_{1} \otimes A_{2})=
  \omega(A_{1} \otimes {\bf 1})\omega({\bf 1} \otimes A_{2}).
  \ee
\end{defin}
In the case of \ca s we require all pure states to be product states.
\begin{defin}
Let \aone\ and \atwo\ be \ca s. We say \ten\ is {\em separated} (in the
$C^{*}$-sense) if and only if:

for all pure states $\omega$ on \ten , $\omega$ is a product state.
\end{defin}

Notice that if $\omega$ is a pure product state of \ten , then
$\omega_{1}:=\omega\mid_{\aone}$ and $\omega_{2}:=\omega\mid_{\atwo}$ are pure
(where \aone\ and \atwo\ are identified with $\aone\otimes{\bf 1}$ and
${\bf 1}\otimes\atwo$, respectively). This follows easily from
Prop.~\ref{Tak413} below.

The structure theorem for separated $C^*$-systems is the following.
\begin{theorem}[Takesaki 1958]\label{Tak414}
Let \aone\ and \atwo\ be \ca s. Then:

$\ten\mbox{ is separated} \Iff \aone \mbox{ or } \atwo \mbox{ is commutative.}$
\end{theorem}
The proof is immediate from Takesaki (1979), Th.~IV 4.14, together with
the above remark.

In the case of \wa s there are
two different notions of separation. Raggio (1981) originally considered
separation with respect to pure states (as we have done
for \ca s). The second definition (Raggio 1983) is more appropriate to the
$W^*$-setting, since it refers only to normal states.
\begin{defin}
Let $\A$ be a \wa , and $\omega$ a normal state on $\A$. $\omega$ is said to be
{\em decomposable} into normal product states if and only if:

$\omega$ lies in the
norm-closure of
the convex hull of the normal product states.
\end{defin}
Intuitively, $\omega$ is a `continuous convex combination' of normal product
states.
\begin{defin}
Let \aone\ and \atwo\ be \wa s.
  \begin{items}
  \item
\wten\ is {\em separated}
{\em with respect to pure states} if and only if:

for all pure states $\omega$ on \wten , $\omega$ is a product state.
  \item
\wten\ is
{\em separated}
{\em with respect to normal states} if and only if:

for all normal states $\omega$ on \wten ,
$\omega$ is decomposable into normal product states.
  \end{items}
    \end{defin}
The second definition needs to be formulated in terms of decomposability,
because pure normal states need not exist in a general $W^*$-algebra.
This makes the two
definitions apparently independent of each other. In both cases, however,
analogues of Takesaki's theorem hold; and the definitions turn out to be
equivalent.

\begin{theorem}[Raggio 1981]\label{Raggio0}
Let \aone\ and \atwo\ be \wa s. Then the following are equivalent:
  \begin{items}
    \item
    $\wten$ is separated w.r.t. pure states;
    \item
    $\aone$ or $\atwo$ is commutative.
  \end{items}
\end{theorem}
\begin{theorem}[Raggio 1983, 1988]\label{Raggio1}
Let \aone\ and \atwo\ be \wa s. Then the following are equivalent:
  \begin{items}
    \item
    $\wten$ is separated w.r.t. normal states;
    \item
    $\aone$ or $\atwo$ is commutative.
  \end{items}
\end{theorem}
The first of these two theorems is proved directly in Raggio (1981). One can
also derive it from Th.~\ref{Tak414}.
For the proof of the second theorem,
see Raggio (1988) and the references therein. (In Raggio (1983) the theorem is
proved under more restrictive assumptions). Since
the two definitions of separation for \wa s coincide, we shall simply talk
of separation in the $W^{*}$-sense.

We now establish the link between the
concept of separation for \ca s and the concept of separation for \wa s via
the representations of a \ca . We shall need it in the proof of Th.~\ref{last}.

\begin{prop}\label{Tak413}
Let \aone\ and \atwo\ be  \ca s, and $\pi_{1}$ and
$\pi_{2}$ two representations
of \aone\ and \atwo , respectively. Then:
  \be
  (\pi_{1}\otimes\pi_{2} (\ten))'' =
  \pi_{1}(\aone)''\:\wotimes\: \pi_{2}(\atwo)''.
  \ee
\end{prop}
This is Takesaki (1979), Prop. IV 4.13.

If $\pi_{1}$ and $\pi_{2}$ are the representations of \aone\ and \atwo\
induced by a representation $\pi$ of \ten, we have
  \be
 \pi ( \ten )=\pi_{1} \otimes \pi_{2} (\ten),
  \ee
and we obtain:
\begin{cor}
\clet , $\pi$ a representation of \ten , and $\pi_{1}$ and $\pi_{2}$ the
induced representations of \aone\ and \atwo . Then
  \be
 \pi ( \ten )'' =
  \pi_{1} (\aone)'' \:\wotimes\: \pi_{2} (\atwo)''.
  \ee
\end{cor}
Thus, the von Neumann algebra associated with a representation of a tensor
product \ca\ has a canonical factorization as a $W^{*}$-tensor product. This
motivates the following definition:
\begin{defin}
\clet , and $\pi$ a representation of \ten .
We say that \ten\ is {\em separated
in the representation $\pi$} if and only if:

$\pi (\aone \otimes {\bf 1})'' \:\wotimes\: \pi ({\bf 1} \otimes \atwo)''$ is
separated (in the $W^{*}$-sense).
\end{defin}

We now prove our first theorem.

\begin{theorem}\label{link}
\clet . The following are equivalent:
  \begin{items}
  \item
    \ten\ is separated;
  \item
    \ten\ is separated in all irreducible representations;
  \item
    \ten\ is separated in the universal representation.
  \end{items}
\end{theorem}
{\sc Proof:}`{\it i)}$\Then${\it ii)}'\\
Take any irreducible representation $\pi$ of \ten . By assumption
\ten\ is separated, so, by Th.~\ref{Tak414}, \aone\ or \atwo\ is
commutative. It follows that $\pi_{1}(\aone)$ or $\pi_{2}(\atwo)$ is
commutative, and so that $\pi_{1}(\aone)''$ or $\pi_{2}(\atwo)''$ is
commutative. By the second half of Th.~\ref{Raggio1} (or by another direct
argument), \ten\ is separated in
the representation $\pi$.

`{\it ii)}$\Leftarrow${\it i)}'\\
Assume \ten\ is not separated. By Th.~\ref{Tak414},
\aone\ and \atwo\ are both noncommutative. Then there exist irreducible
representations  $(\pi_{1},\H_{1})$ of \aone\ and $(\pi_{2},\H_{2})$ of
\atwo , with $\dim\H_{1}, \dim\H_{2} \geq 2$. Irreducibility means that
  \be
\pi_{1}(\aone)'' = \B(\H_{1}), \mbox{ and }
\pi_{2}(\atwo)'' = \B(\H_{2}).
  \ee
But $\B(\H{_1})$ and $\B(\H{_2})$ are both
noncommutative, and it follows by the first half of Th.~\ref{Raggio1} that
  \be
  \pi_{1}(\aone)'' \:\wotimes\: \pi_{2}(\atwo)''
  \ee
is not separated. Consequently,
\ten\ is not separated in the representation $\pi:=\pi_1\otimes\pi_2$.

`{\it i)}$\Iff${\it iii)}'\\
Let $\pi$ be the universal representation of \ten .
By Th.~\ref{Tak414}, \ten\ is separated iff $\A_1$ or $\A_2$ is
commutative. This is the case iff $\pi_1(\A_1)$ or $\pi_2(\A_2)$ is
commutative, iff $\pi_1(\A_1)''$ or $\pi_2(\A_2)''$ is commutative. By
Th.~\ref{Raggio1},
this is equivalent to separation in the representation $\pi$.
\hspace{2em}\rule{1ex}{1ex}

\section{Bell inequalities and separation}\label{Bell}
Separation means that two systems never present EPR-correlations, in other
words, that any possible correlations between the two can be explained in terms
of classical mixtures. It is well-known that the problem of whether given sets
of numbers that look like probabilities and correlations can be fitted into a
classical probabilistic scheme is related to the Bell inequalities. Results of
this kind are contained in particular in Pitowsky (1989); see also Fine (1982
and references therein)
and Prof.\ Beltrametti's contribution to the present volume. It is thus natural
to ask whether there is a way of characterising separation using Bell
inequalities. In fact there is: Baez (1987) obtained some partial results for
$C^*$-algebras, and Raggio (1988) showed that in the case of \wa s, a composite
system \wten\ is separated if and only if all normal states on \wten\ satisfy
the Bell inequalities (in the sense of Def.~\ref{ineq} below). We prove the
analogous results for \ca s.

In 1987, Baez published a short note on Bell inequalities in the \ca ic
formalism. His main results are the following.

\begin{defin}\label{ineq}
Let \aone\ and \atwo\ be $C^{*}$- ($W^{*}$-) algebras.
We say a state $\omega$ on \ten\ (\wten )
{\em satisfies the Bell inequalities} iff \mbox{$\fa A,A'\in\aone$},
\mbox{$\fa B,B'\in \atwo$} with norm less or equal to $1$,
  \be
  \abs{\omega (A \otimes (B-B'))} + \abs{\omega (A' \otimes (B+B'))} \leq 2.
  \ee
\end{defin}
\begin{prop}\label{Baez1}
Let \aone\ and \atwo\ be \ca s, and let $\omega$ be a product state
on \ten . Then $\omega$ satisfies
the Bell inequalities.
\end{prop}
The proof is completely analogous to the standard proofs of the Bell
inequalities: see Baez (1987).

\begin{defin}\label{nothv}
Let $\A$ be a \ca , and let $\omega$ be a state on $\A$. $\omega$ is said to be
{\em decomposable} into product states if and only if:

$\omega$ lies in the weak$^{*}$-closure of
the convex hull of the product states.
\end{defin}

\begin{prop}\label{decom}
Let \aone\ and \atwo\ be \ca s, and let \ten\ be separated. Then:

for all states $\omega$ on \ten ,
$\omega$ is decomposable into product states.
\end{prop}
{\sc Proof:} By the Krein-Milman theorem, every state on \ten\ lies in the
weak$^{*}$-closure of the convex hull of the pure
states. By the assumption of separation, these are product
states.\hspace{2em}\rule{1ex}{1ex}

From this and Th.~\ref{Tak414} Baez obtains:
\begin{theorem}[Baez 1987]\label{Baez2}
Let \aone\ and \atwo\ be two \ca s, and let \aone\ or \atwo\ be commutative.
Then:

for all states $\omega$ on \ten , $\omega$ satisfies the Bell inequalities.
\end{theorem}
{\sc Proof:}
For a product state, the theorem reduces to
Prop.~\ref{Baez1}. For an arbitrary
state, it follows by convexity
and continuity using Prop.~\ref{decom}.

Baez states explicitly that for algebras of type I also the converse holds,
because in this case one can reproduce the Bohm-Bell
situation, which gives violation of the Bell inequalities. We shall now prove 
more generally that for arbitrary $C^{*}$-algebras $\A_1$ and $\A_2$, the product \ten\ is
separated {\em if and only if} $\A_1$ or $\A_2$ is commutative. This is the
analogue of Raggio's
results for $W^*$-algebras (Th.~\ref{Raggio2}),
from which in fact our theorem follows.
\begin{theorem}[Raggio 1988]\label{Raggio2}
\wlet . Then the following are equivalent:
  \begin{items}
    \item
    \aone\ or \atwo\ is commutative;
    \item
    for all normal states $\omega$ on \wten ,
    $\omega$ satisfies the Bell inequalities.
  \end{items}
\end{theorem}
For the proof, see Raggio (1988). The trick is to construct a Bohm-Bell type
counterexample not using finite-dimensional projections, which might not exist
(type III), but constructing subalgebras isomorphic to the two-by-two matrices
using pairs of generally {\em infinite}-dimensional noncommuting projections.

Finally, we prove the converse of
Prop.~\ref{decom} and Th.~\ref{Baez2}.

\begin{theorem}\label{last}
\clet . The following are equivalent:
  \begin{items}
    \item
      \aone\ or \atwo\ is commutative;
    \item
      \ten\ is separated;
    \item
      for all states $\omega$ on \ten , $\omega$ is decomposable into product
      states;
    \item
      for all states $\omega$ on \ten ,
      $\omega$ satisfies the Bell
      inequalities.
  \end{items}
\end{theorem}
{\sc Proof:}
We will prove that if all $\omega$ satisfy the Bell inequalities,
\ten\ is separated
in all $\pi$ (the proof using the universal representation is analogous). From
Th.~\ref{link}, the theorem will follow.
Take an irreducible representation $(\pi,\H)$
of \ten , and any $\psi$ in
$\H$. $\psi$ is a state on $\pi(\ten)''$ and defines a (pure)
state on \ten\ via
  \be
  \omega_{\psi} (A) := (\psi, \pi (A)\psi).
  \ee
By assumption, $\omega_{\psi}$ satisfies the Bell inequalities, \ie
\mbox{$\fa A,A'\in\aone$}, \mbox{$\fa B,B'\in\atwo$}
with norm less or equal to 1,
  \be
  \begin{array}{cl}
     &   \abs{(\psi, \pi(A \otimes (B-B')) \psi)} +
         \abs{(\psi, \pi(A'\otimes (B+B')) \psi)} =    \\
  =  &   \abs{(\psi, \pi(A) \otimes (\pi(B)-\pi(B')) \psi)}  +
         \abs{(\psi, \pi(A')\otimes (\pi(B)+\pi(B')) \psi)}   \leq  2.
    \end{array}
  \ee
By continuity, we also have \mbox{$\fa A,A'\in\pi(\aone)''$},
\mbox{$\fa B,B'\in\pi(\atwo)''$} with norm less or equal to 1,
  \be
  \abs{(\psi, A  \otimes (B-B') \psi)}  +
  \abs{(\psi, A' \otimes (B+B') \psi)}  \leq  2.
  \ee
Thus, $\psi$ as a state on $\pi(\ten)''$
satisfies the Bell inequalities. But now,
any normal state on
$\pi(\ten)''$ lies in the norm-closure of the convex hull of states of the form
$\psi$. So, by convexity
and continuity, it satisfies the Bell inequalities.

By Th.~\ref{Raggio2}, $\pi(\aone)'' \:\wotimes\: \pi(\atwo)''$ is separated,
\ie \ten\ is separated in the representation $\pi$.
Since $\pi$ was arbitrary, \ten\ is separated in all irreducible
representations.

By Th.~\ref{link}, \ten\ is separated.
\hspace{2em}\rule{1ex}{1ex}

\section{For whom the Bell tolls}\label{conc}
The original derivations of the Bell inequalities by Bell and others proceeded
from a hypothesis
of local hidden variables that corresponds to a classical (albeit generally
indeterministic and certainly
contextual) model for the probabilities to be expected in
EPR-type experiments. Classical examples of composite systems also trivially
satisfy the Bell inequalities. `Classicality implies the Bell inequalities'
seems to be the conclusion. The converse also appears to be true since the
violation of the Bell inequalities by quantum mechanical systems is what makes
them interesting in the first place as a touchstone for testing local hidden
variable theories experimentally.

One sense in which the converse has been, indeed,  shown to hold rigorously is
Pitowsky's
analysis (1989) of the Bell inequalities: satisfaction of the Bell
inequalities is a necessary and sufficient condition for the existence of a
classical probability space and probability distributions reproducing the
given data. This is the same situation as we have here:
Theorems \ref{Raggio2} and \ref{last} indeed
show that all possible correlations between results of measurements on the
systems represented by $\A_{1}$ and $\A_{2}$ are purely classical.

However, Pitowsky's results are not vindicating any correspondence between
classicality and the Bell inequalities other than in the precise form he gave:
as a matter of fact,
{\em it is just the correlations that are classical}. Indeed, the catch-phrase
`Bell inequalities mean that everything is classical' is wrong. Theorems
\ref{Raggio2} and \ref{last} show that, contrary to the case of traditional
Hilbert-space quantum mechanics, in algebraic quantum mechanics there are
systems that satisfy the Bell inequalities in all possible states, and
nevertheless are {\em not} entirely classical --- indeed, one of the subsystems
may be completely quantum mechanical.

Neither is it the case that the system, even though not classical, is described
by `classical' hidden variables. In a hidden
variable account, one
tries to explain the EPR-correlations present in traditional Hilbert-space
quantum mechanics by adding `hidden variables' to
the theory. Separation in our sense requires instead that we limit ourselves
to systems that are {\em not} EPR-correlated, thus remaining {\em within}
quantum mechanics, albeit within
the more general formalism of algebraic
quantum mechanics.

Thus classicity of behaviour in
the sense of the Bell inequalities being satisfied
is not sufficient for
the system as a whole to be classical --- only a subsystem is. Further,
Aerts (1991) has argued that the Bell inequalities are not even
{\em necessary} for a system to be classical: even a classical system can
violate the Bell
inequalities if it cannot
be separated
into individual subsystems. The relation between separation and the Bell
inequalities seems to suggest that what lies physically --- or metaphysically
--- behind the Bell
inequalities (and the mathematical existence of classical probabilities) is the
{\em individuality} of the systems considered: each system has a physical state
of its own, uninfluenced by the states of other systems, a special instance of
this being Bell's original concept of locality for hidden variables.

\section*{Acknowledgements}
I wish to thank first of all Jeremy Butterfield for the endless and immensely
helpful discussions I had with him. My thanks go further to the audiences who
in some form or other heard and commented on this material. In particular I
wish to thank Anton Amann, Harvey Brown, Fiona Harrison, Klaas
Landsman, and Constantin Piron. [I am also very grateful to Adrian Kent for 
alerting me to the inaccuracy of some side remarks in the published version, 
which have been accordingly edited out of this online version.] 
This work was supported by the British Academy
and the Arnold Gerstenberg Fund.

\newpage

\section*{References}
Aerts, D. (1991): `A mechanistic classical laboratory situation violating the
Bell inequalities with $2\cdot\sqrt{2}$, exactly ``in the same way'' as its
violations by the EPR experiments' {\em Helv. Phys. Acta}, {\bf 64},
\mbox{1-23}.

Baez, J. (1987): `Bell's inequality for \ca s', {\em Lett. Math. Phys.},
{\bf 13}, \mbox{135-136}.


Fine, A. (1982): `Joint distributions, quantum correlations, and commuting
observables', {\em J. Math. Phys.}, {\bf 23}, \mbox{1306-1310}.

Pitowsky, I. (1989): {\em Quantum Probability --- Quantum Logic}. Springer,
Berlin and London.

Primas, H. (1978): {\em Quantenmechanik und Theoretische Chemie}. Lecture
course, ETH Z\"{u}rich, Summer Term 1978.

Primas, H. (1983): {\it Chemistry, Quantum Mechanics and
Reductionism}. 2nd edition. Springer, Berlin.

Raggio, G.A. (1981): {\em States and Composite Systems in $W^{*}$-algebraic
Quantum Mechanics}, Dissertation ETH No. 6824, Z\"{u}rich.

Raggio, G.A. (1983): {\em Zusammengesetzte Systeme in der verallgemeinerten
$W^*$-algebraischen Quantenmechanik}. Lecture, Karl-Marx-Universit\"{a}t,
Leip\-zig, November 1983.

Raggio, G.A. (1988): `A remark on Bell's inequality and decomposable normal
states', {\em Lett. Math. Phys.}, {\bf 15}, \mbox{27-29}.


Takesaki, M. (1979): {\em Theory of Operator Algebras I}. Springer, New York.

\end{document}